# Numerical Simulation of $Cs_2AgBiBr_6$-based Perovskite Solar Cell with ZnO Nanorod and P3HT as the Charge Transport Layers

*Intekhab Alam[1*], Rahat Mollick[1], Md Ali Ashraf[2]*

[1]Department of Mechanical Engineering (ME), Bangladesh University of Engineering and Technology (BUET), East Campus, Dhaka-1000, Bangladesh

[2]Department of Industrial and Production Engineering (IPE), Bangladesh University of Engineering and Technology (BUET), East Campus, Dhaka-1000, Bangladesh

*Corresponding Author
Email: intekhabsanglap@gmail.com, Phone No: +8801819279881

**Abstract**

We carried out simulative investigations on a non-toxic, lead-free perovskite solar cell (PSC), where $Cs_2AgBiBr_6$, P3HT, ZnO nanorod, and C were utilized as the absorber layer, hole transport layer, electron transport layer, and back contact, respectively. At 600 nm optimum absorber thickness, the device achieved a maximum power conversion efficiency of 4.48%. The PSC operated optimally when the electron affinities were set at 3.3 eV and 4.6 eV for P3HT and ZnO nanorod, respectively. Moreover, the hole mobility and acceptor concentration of P3HT should be weighed during the choosing of appropriate doping additives and doping levels. Besides, the optimum back contact work function and absorber defect density were found to be 5.2 eV and $10^{15}$ cm$^{-3}$, respectively. We also observed the effect of radiative recombination rates and different charge transport layers on the device's performance. Overall, this study's simulation results will provide insightful guidance towards fabricating an environmentally benign PSC.

# 1. Introduction

The world is proceeding towards clean, renewable power with the calls for phasing out fossil fuel-driven energy conversion technologies.[1] Among the available renewable sources of energy, solar power is the most plentiful.[2] For exploiting this huge amount of power, the use of photovoltaic (PV) solar cells is on the rise. Their utilization will certainly play a noteworthy role in future sustainable energy systems.[3] The most commonly used Si-based PV solar cells have some major shortcomings such as low efficiency and high cost.[4,5] Therefore, other emerging PV technologies such as dye-sensitized solar cells (DSSCs), organic photovoltaic (OPV), quantum dots solar cells (QDSCs), hybrid organic-inorganic solar cells, and perovskite solar cells (PSCs) have attracted the researchers due to their promising performance and low production costs.[5,6] Moreover, the incorporation of perovskite material into DSSC as the sensitizer layer hugely expedited the third-generation solar cell advancement.[7]

Organic-inorganic halide perovskites have the form of $AMX_3$, where A is a monovalent cation, M is a divalent metal, and X is a halide anion. They have gathered attention in the field of PV technology due to their high absorption coefficient, large electron and hole diffusion length, career lifetime, and good charge carrier mobility.[8,9] The power conversion efficiency (PCE) of the most commonly used PSCs based on organic-inorganic halide $CH_3NH_3PbI_3$ perovskite has enhanced from 3.8% in 2009 to a record 23.3% in less than a decade, surpassing the performance of traditional CdTe and $Cu(In, Ga)Se_2$-based thin-film solar cells.[10,11] Furthermore, the tandem solar cells based on metal halide perovskites have obtained more than 26% PCE, thus competing with the crystalline silicon (c-Si) in performance.[12,13] Although the optical and electrical properties of Pb-based perovskites facilitate high PCE, their commercial application is hindered by their instability towards heat, oxygen, and moisture along with the toxic nature of lead.[14,15] Human exposure to lead can cause premature infants and low birth rates, circulatory system damage, and brain and kidney malfunction.[7] These obstacles paved the way for the researchers to focus on Pb-free, non-toxic replacement of $CH_3NH_3PbI_3$. In our previous work, we utilized non-toxic, Pb-free $CH_3NH_3SnI_3$ perovskite as the alternative.[16] However, it is highly susceptible to the atmosphere due to its facile oxidation that results in low PCE.[16,17,18] On the contrary, halide double perovskites with 3-D structures having the form of $A_2M'M''X_6$, where A is a monovalent cation, M' and M'' are monovalent and trivalent metal ions, respectively, and X is a halide anion, have become promising alternatives. Because of their high environmental stability, encouraging optoelectronic properties, and low toxicity, they can be considered as promising replacements for Pb-based perovskites.[19,20,21] Especially, lead-free, non-toxic, Bi-based $Cs_2AgBiBr_6$ halide double perovskite is a good candidate for PV solar cell. High thermal and ambient stability, long carrier recombination lifetime, high absorption coefficient, and proper bandgap of $Cs_2AgBiBr_6$ perovskite are the reasons behind its suitability as the absorber layer.[22,23,24] Besides, this perovskite crystallizes in the elpasolite structure, absorbs light in the visible range of the solar spectrum, and has an indirect bandgap ranging from 1.83 eV to 2.19 eV.[25,26] The bandgap engineering operated by defect inducing and trivalent metal alloying along with the device optimization exhibit great potentiality in improving the 7.92% theoretical PCE achieved by the $Cs_2AgBiBr_6$-based PSC.[27,28]

Electrical and optical properties, stability, and cost-effectiveness are the main considering factors while choosing a hole transport layer (HTL) for the PSCs.[5] Among the commonly used organic

HTLs, 2,2',7,7'-tetrakis[N,N-di(4-methoxyphenyl)amino]-9,9'-spirobifluorene (spiro-OMeTAD), poly(3-hexylthiophene-2,5-diyl) (P3HT), and poly(3,4-ethylenedioxythiophene): poly(styrenesulfonatuze) sulfonic acid (PEDOT: PSS) are a few.[29] A series of p-type inorganic metal compounds including $CuO$, $Cu_2O$, and $MoO_3$ have also been employed in the PSCs as the HTL.[30] Although highly efficient and stable PSCs commonly use inorganic HTLs, the organic HTLs facilitate superior processability and tunability, making their use more favorable.[31] Organic Spiro-OMeTAD is the most widely used HTL for PSCs. But it has very low hole mobility and poor conductivity attributed to its triangular pyramid structure and large intermolecular distance.[32,33] Moreover, its hole mobility and conductivity are contingent on the dopants, which can deteriorate the device's stability.[34] Its commercial application is hindered because of its high fabrication cost along with its thermal and chemical instability.[35,36] Contrarily, undoped P3HT HTL has proven to be extremely effective in achieving high PCE.[37] In addition, a robust hydrophobic HTL like P3HT instead of hydrophilic Spiro-OMeTAD can considerably improve the PSCs' moisture stability.[31] Low-cost (10 times cheaper than spiro-OMeTAD) P3HT possesses unique optoelectronic properties such as high hole mobility, wide bandgap, good solubility, and thermal stability.[36,37,38] Besides, by altering the length of the side chains, molecular weight, and temperature and using the directional crystallization techniques, its electrical, optical, and transport properties can be easily tuned.[36,39] Furthermore, it has shown quite remarkable results to date, retaining 80% of the first efficiency after 1008 hours under 85% relative humidity at the room temperature.[29]

The electron transport layer (ETL) affects the PV performance of PSCs by dominating the charge carrier extraction, transportation, and recombination process. Therefore, it is crucial to choose an ETL with proper band alignment, high electron mobility, adequate light transmittance, and moisture resistance.[40,41,42] Inorganic n-type metal oxides such as $TiO_2$ and ZnO have been favorable for highly efficient and stable PSCs.[43] Since 2009, $TiO_2$ has been widely used as the ETL for PSCs due to its favorable conduction band position, wide bandgap, and long carrier lifetime.[44] However, it has several limitations due to its high-temperature fabrication, intrinsic low electron mobility, and UV-light instability under sustained illumination.[45,46] Meanwhile, ZnO is a suitable replacement of $TiO_2$ as the ETL because of its excellent transparency in the visible range, high electron mobility, large bandgap (~3.4 eV), high thermal and electrochemical stability, low fabrication cost, and excellent optoelectronic properties.[40,47,48] Various morphological structures of ZnO such as thin-film, single crystal, nanowire, and nanorod have been synthesized using the low-temperature solution processes.[49,50] According to the published literature, ZnO nanorods (ZnO-NRs) have been utilized in the PSC to efficiently collect the electron from the lower wavelength region.[51,52] Furthermore, due to the faster electron mobility, higher conduction band, and higher electron density than the pure ZnO, Al-doped ZnO/ ZnO-NR can improve the performance of PSC.[53] Besides, vertically aligned ZnO-NRs can enhance the donor/ acceptor interface, and they can form direct pathways towards the electrode with high electron mobility for electron transport.[54,55]

Transparent conducting oxides (TCOs) like fluorine-doped tin oxide (FTO) and tin-doped indium oxide (ITO) have found numerous applications in the PSCs.[56] But indium is more costly compared to fluorine and tin, making the ITO highly expensive.[57] In addition, tunable bandgap, good electrical conductivity, and chemical stability highlight FTO as an excellent choice for the PSCs and a suitable alternative to the ITO.[56,58] Furthermore, over the years, most of the PSCs have been fabricated using noble metal electrodes such as gold and silver as the back contact.[8] The high energy-consuming vacuum thermal evaporation process of these back contacts raises the cost of the large-scale production of PSCs.[59] Therefore, earth-abundant, cheap, anti-corrosive, water-proof, and easy-to-synthesized carbon has been studied as the alternative back contact.[60,61] Also, the high conductivity and good stability of the carbon back contacts make their use in the PSCs even more advantageous.[62]

The sequential vapor deposition method was employed previously to experimentally fabricate the FTO/ c-TiO$_2$/ Cs$_2$AgBiBr$_6$/ P3HT/ Au solar cell with a maximum PCE of 1.37%.[63] Besides, for the experimental fabrication of ITO/ SnO$_2$/ Cs$_2$AgBiBr$_6$/ P3HT/ Au solar cell with a maximum PCE of 1.44%, the low-pressure assisted solution process was utilized.[27] In this simulation, we proposed a novel device having an organic-inorganic structure of soda-lime glass (SLG)/ FTO/ ZnO-NR/ Cs$_2$AgBiBr$_6$/ P3HT/ C. We observed the effect of absorber thickness, the acceptor density and hole mobility of HTL, band-to-band radiative recombination rate, the electron affinity of HTL and ETL, back contact work function, absorber defect density, and alternative charge transport layers on the PSC performance. All these simulation results will assist to experimentally fabricate a lead-free, non-toxic, and viable halide double PSC.

## 2. Methodology

SCAPS 1-D (a Solar Cell Capacitance Simulator) was utilized to carry out the computations. The program was developed at the Department of Electronics and Information Systems of the University of Gent, Belgium.[64] It is written in the C programming language. Various profiles, including grading, generation, recombination, and defects of device architecture, can be calculated using this simulation program.[65] SCAPS 1-D has already been employed for the simulation of PSCs in the published literature.[16,66] The software numerically solves the Poisson's and continuity equations to calculate parameters such as open-circuit voltage ($V_{OC}$), photogenerated current density ($J_{SC}$), fill factor (FF), and PCE.[16,67] The Poisson's equation (equation 1), electron continuity equation (equation 2), and hole continuity equation (equation 3) are given below:[68]

$$\frac{d}{dx}\left(-\varepsilon(x)\frac{d\psi}{dx}\right) = q[p(x) - n(x) + N_d^+(x) - N_a^-(x)] \ldots \ldots \ldots (1)$$

$$\frac{\partial j_n}{\partial x} = q\left(R_n - G + \frac{\partial n}{\partial t}\right) \ldots \ldots \ldots (2)$$

$$\frac{\partial j_p}{\partial x} = -q\left(R_p - G + \frac{\partial p}{\partial t}\right) \ldots \ldots \ldots (3)$$

Where, $\varepsilon$ is the permittivity, $q$ is the electron charge, $\psi$ is the electrostatic potential, $n$ is the total electron density, $p$ is the total hole density, $N_d^+$ is the ionized donor-like doping concentration, $N_a^-$ is the ionized acceptor-like doping concentration, $j_n$ and $j_p$ are the electron and hole current densities, respectively, $R_n$ and $R_p$ are the net recombination rates for electron and hole per unit volume, respectively, and $G$ is the generation rate per unit volume.

To validate the simulation results found from the SCAPS, the simulations were also carried out using wxAMPS. wxAMPS is a free simulation software package written in C++ that was developed by the University of Illinois at Urbana Champaign. It is based on the original AMPS (Analysis of Microelectronic and Photonic Structures) code developed at the Pennsylvania State University. wxAMPS incorporates two different tunneling models for better simulation of specific solar cells and has the versatility to incorporate multi-junction device modeling. It numerically solves the Poisson's and continuity equations required for charge carrier transport, and the equations involved along with the comparative results have been illustrated in the Supplementary File.[65,69,70] It was found that there were almost similar $V_{OC}$, $J_{SC}$, FF, and PCE trends during the absorber layer thickness optimization study. However, there were differences in their values, and also, the optimum absorber thickness found from the wxAMPS was lower than the one found from the SCAPS. Moreover, the maximum PCE found from the wxAMPS was 0.49% higher than the SCAPS in magnitude.

**2.1 Architecture of the Devices**

The structure of the SLG/ FTO/ ZnO-NR/ Cs$_2$AgBiBr$_6$/ P3HT/ C solar cell has been illustrated in Figure 1.[63,71,72] P3HT was utilized as the HTL over the C back contact, Cs$_2$AgBiBr$_6$ was the absorber layer, ZnO-NR was the ETL, and FTO was the TCO. It was a solid-state planar heterojunction n-i-p halide double PSC with the Cs$_2$AgBiBr$_6$ absorber sandwiched between the ZnO-NR ETL and P3HT HTL. SLG was positioned over FTO, and the PSC absorbed the sunlight from illumination over the SLG side.

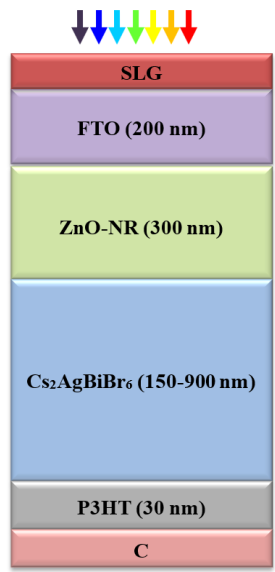

**Figure 1**: Device architecture of the proposed PSC.

## 2.2 Simulated parameters

The simulations were carried out using material parameters that were cautiously selected from the published literature.[71-76] These material parameters have been listed in table 1. The temperature was taken as 300K with the standard illumination of AM1.5G. The electron and hole thermal velocities were kept constant at $10^7$ cms$^{-1}$ for each layer. The work function for the C back contact was 5 eV, and thermionic emission/ surface recombination velocities for electrons and holes were $10^5$ cms$^{-1}$ and $10^7$ cms$^{-1}$, respectively.[61] For each layer, the optical absorption constant, α(hv), was set by the new "Eg-sqrt" model, and the model details have been shown in the Supplementary File.[64,65,77]

**Table 1**: Input material parameters used in the PSC simulation.

| | Units | TCO | ETL | Absorber Layer | HTL |
|---|---|---|---|---|---|
| Material | | FTO | ZnO-NR | Cs$_2$AgBiBr$_6$ | P3HT |
| Thickness | nm | 200 | 300 | 150-900 | 30 |
| Bandgap ($E_g$) | eV | 3.20 | 3.47 | 2.05 | 2.00 |
| Electron affinity (χ) | eV | 4.40 | 4.30 | 4.19 | 3.20 |
| Relative Permittivity ($ε_r$) | - | 9.00 | 9.00 | 5.80 | 3.00 |
| CB effective density of states ($N_c$) | cm$^{-3}$ | 2.2E+18 | 2.0E+18 | 1.0E+20 | 1.0E+20 |
| VB effective density of states ($N_v$) | cm$^{-3}$ | 1.8E+19 | 1.8E+20 | 1.0E+20 | 1.0E+20 |
| Electron mobility ($μ_n$) | cm$^2$V$^{-1}$s$^{-1}$ | 20 | 100 | 11.81 | 1.0E-4 |
| Hole mobility ($μ_p$) | cm$^2$V$^{-1}$s$^{-1}$ | 10 | 25 | 0.49 | 1.0E-4 |
| Donor density ($N_d$) | cm$^{-3}$ | 1.0E+18 | 1.0E+19 | 1.0E+19 | 0 |
| Acceptor density ($N_a$) | cm$^{-3}$ | 0 | 0 | 1.0E+19 | 1.0E+16 |
| Radiative Recombination | cm$^3$s$^{-1}$ | 2.3E-9 | 2.3E-9 | 2.3E-9 | 2.3E-9 |

## 3. Result and Discussion

### 3.1 Effect of the Absorber Layer Thickness

All of the layers were considered defect-free primarily to obtain the theoretical maximum of the device performance that can be approached from an experimental point of view.[78] The absorber layer thickness is a cardinal component to determine the PCE of a PSC. Because very low absorber thickness causes a drop in the absorption rate and PCE, whereas very high absorber thickness creates an impediment in the charge carrier's movement to the charge collecting layers. However, a PSC generally obtains very high values of $J_{SC}$ and PCE at a low absorber thickness due to the very high absorption coefficient of the perovskite material.[79] For determining the optimum absorber thickness, the thickness of the Cs$_2$AgBiBr$_6$ absorber was varied between 150 nm to 900 nm, while other material parameters were kept invariable. Critical parameters such as PCE, FF, $V_{OC}$, and $J_{SC}$ were simulated against the varying absorber thickness, and the observed trends have been exhibited in figure 2. In addition, the quantum efficiency (QE) vs. wavelength curve for the optimum absorber thickness of PSC has been illustrated in figure 3.

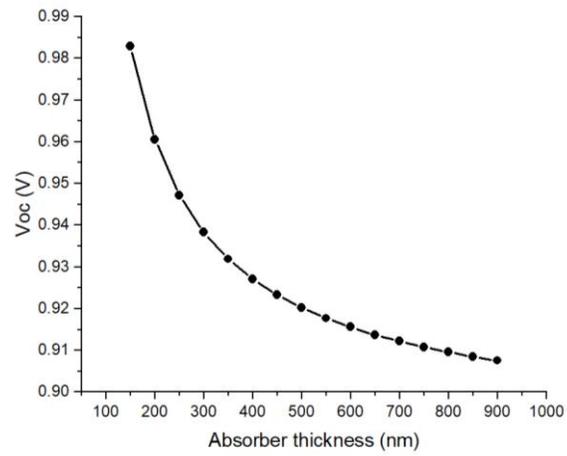

**(a)**

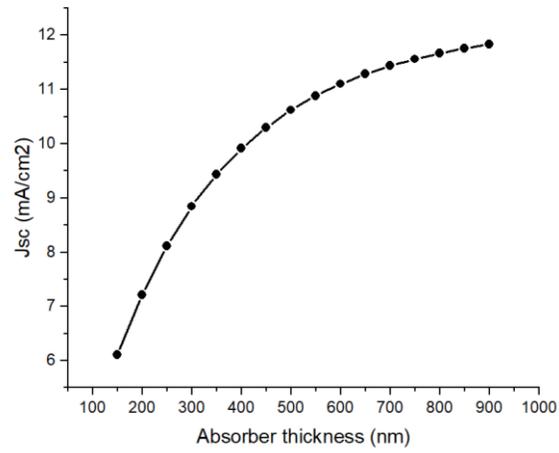

**(b)**

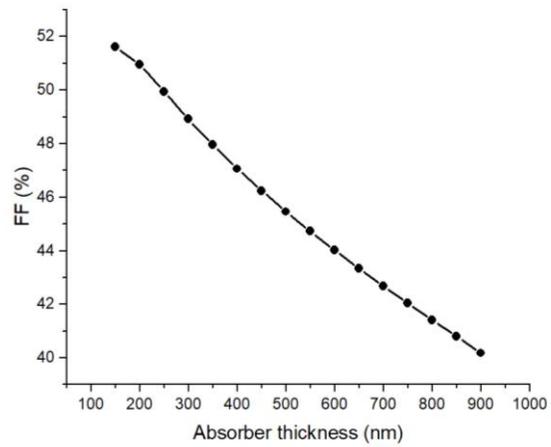

**(c)**

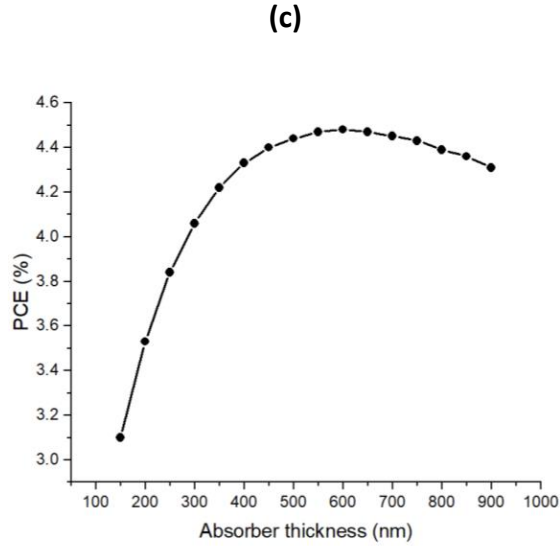

**(d)**

**Figure 2**: **(a)** $V_{OC}$, **(b)** $J_{SC}$, **(c)** FF, and **(d)** PCE diagram with varying absorber thickness.

The maximum PCE of 4.48% ($J_{SC}$ = 11.104 mA/cm$^2$, FF = 44.02%, and $V_{OC}$ = 0.9156 V) was found at 600 nm absorber thickness, which was taken as the optimum thickness. The PCE increased up to the optimum absorber thickness, and it started to decline after reaching the maximum value. Moreover, the optimum thickness of $Cs_2AgBiBr_6$ perovskite was similar to the published literature.[68] When the absorber thickness was increased from 150 nm to 900 nm, the $J_{SC}$ increased steadily from 6.1067 mA/cm$^2$ to 11.8308 mA/cm$^2$, while the FF and $V_{OC}$ decreased steadily from 51.62% to 40.18% and 0.9829 V to 0.9076 V, respectively. A thicker absorber layer absorbs more photons that results in the generation of more electron-hole pairs. This is the reason behind the gradual rise of $J_{SC}$ when plotted against the increasing absorber thickness. However, the $J_{SC}$ declined at a faster rate when the absorber thickness was less than 250 nm. It was because, at this range, the recombination enhances near the C back contact, causing an enhanced electron-hole pair reduction. The following equation can be utilized to understand the $V_{OC}$ kinetics:

$$V_{OC} = \frac{AK_BT}{q}\left[ln\left(1+\frac{I_l}{I_0}\right)\right] \ldots \ldots \ldots (4)$$

Where, $V_{OC}$ is the open-circuit voltage, $A$ is the ideality factor, q is an elementary charge, $\frac{K_BT}{q}$ is the thermal voltage, $I_l$ is the light generated current, and $I_0$ is the dark saturation current. Equation (4) identifies the inverse relationship between the $I_0$ and $V_{OC}$. The $I_0$ increases with the increasing absorber thickness that causes a drop in $V_{OC}$.[16,71] Furthermore, lower free hole concentration in the thicker absorber at the open-circuit can cause a reduction in $V_{OC}$.[80] Meanwhile, the FF is mainly a function of resistive losses.[81] The charge pathway resistance increases with the increasing absorber thickness, leading to a decrease in FF. Finally, the PCE is a function of $J_{SC}$, FF, and $V_{OC}$. Increasing the absorber thickness enhances the electron-hole pair generation through more absorption of light, causing an initial spike in PCE. However, concurrently, the possibilities of radiative recombination and charge pathway resistance enhance in the thicker absorber layer. This causes electron-hole pair reduction, which in turn, reduces the PCE.[68]

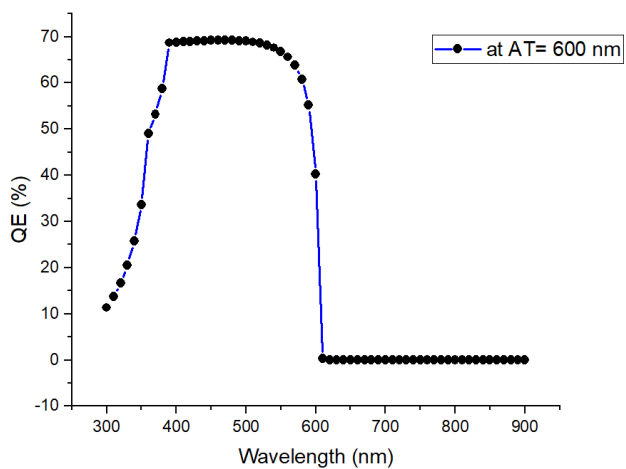

**Figure 3**: QE vs. wavelength diagram for the optimum absorber thickness (AT).

## 3.2 Effect of the Acceptor Density ($N_a$) and Hole Mobility ($\mu_p$) of HTL

The acceptor density ($N_a$) and hole mobility ($\mu_p$) of P3HT were modulated to understand their impact on the PSC performance. Firstly, the $N_a$ of P3HT was varied from $10^{15}$ cm$^{-3}$ to $10^{19}$ cm$^{-3}$ for the best absorber thickness, and the variation in PCE, FF, $V_{OC}$, and $J_{SC}$ has been captured in figure 4. In addition, the alteration in the total recombination profile of the PSC with different $N_a$ of P3HT has been shown in figure 5.

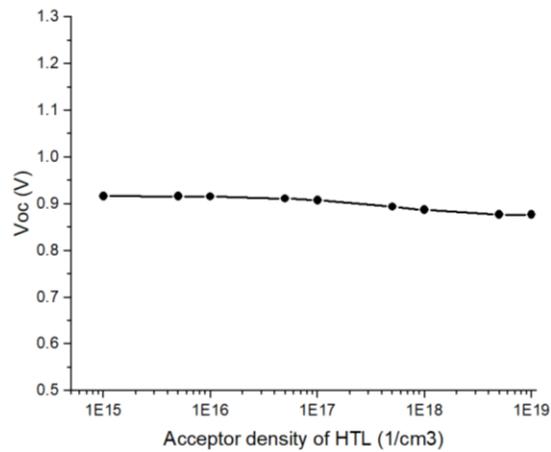

(a)

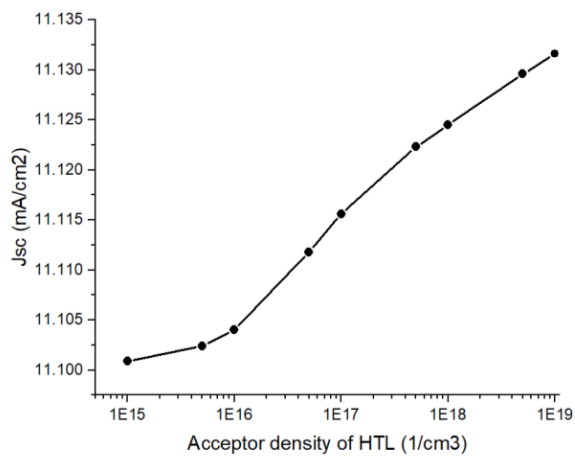

(b)

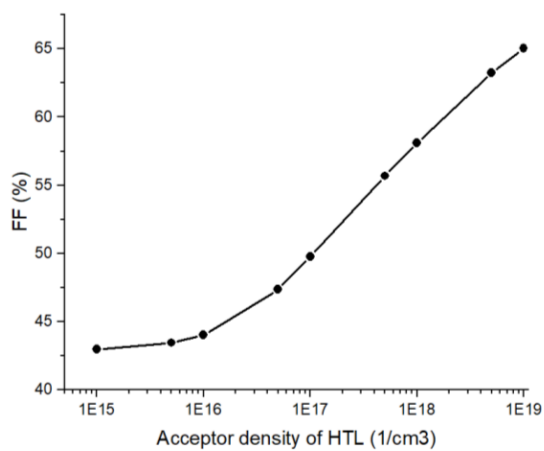

(c)

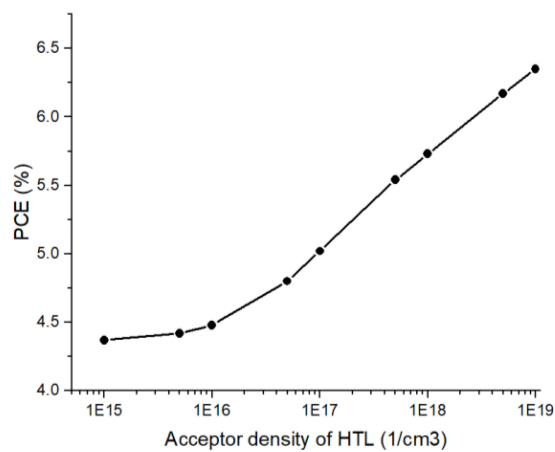

(d)

**Figure 4**: **(a)** $V_{OC}$, **(b)** $J_{SC}$, **(c)** FF, and **(d)** PCE diagram with varying $N_a$ of P3HT for the best absorber thickness.

At the base $\mu_p = 10^{-4}$ cm²V⁻¹s⁻¹ of P3HT, the maximum PCE of 6.35% ($J_{SC}$ = 11.1316 mA/cm², FF = 65.02%, and $V_{OC}$ = 0.8777 V) was attained at $10^{19}$ cm⁻³ $N_a$ of P3HT for the best absorber thickness. The PCE, FF, and $J_{SC}$ of the PSC followed an upward trend when plotted against the increasing $N_a$ of P3HT, whereas the $V_{OC}$ followed a downward trend. The observed trends were similar to the published literature.[82] The PCE and FF improved significantly, but the alterations in $J_{SC}$ and $V_{OC}$ were marginal. When the $N_a$ of P3HT was increased from $10^{15}$ cm$^{-3}$ to $10^{19}$ cm$^{-3}$, the PCE and FF improved from 4.37% to 6.35% and 42.98% to 65.02%, respectively. However, there was only 0.0307 mA/cm$^2$ improvement in $J_{SC}$ and only 0.039 V declination in $V_{OC}$, suggesting that the $N_a$ of P3HT had a negligible impact on them. The PCE was low at the lower value of $N_a$ because of the high series resistance.[83,84] Incorporating p-type dopants in the HTL improves the device performance remarkably as they enhance the $\mu_p$ and charge density of HTL, respectively or simultaneously. This, in turn, improves the conductivity of HTL.[79] That's why, with the augmentation of doping concentration or $N_a$ of P3HT, the PCE improved significantly. The drastic improvement of FF could be due to a drop in series resistance and an increase in shunt resistance with the increasing $N_a$ of HTL.[85]

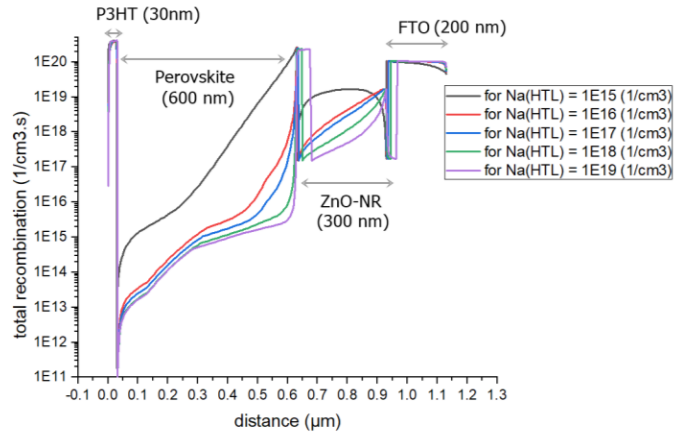

**Figure 5**: Variation in total recombination profile of the PSC with different $N_a$ of P3HT.

Furthermore, for understanding the correlation between the $N_a$ and $\mu_p$ of HTL, the $\mu_p$ of P3HT was enhanced from $10^{-5}$ cm$^2$V$^{-1}$s$^{-1}$ to 10 cm$^2$V$^{-1}$s$^{-1}$ for $N_a = 10^{16}$ cm$^{-3}$, $N_a = 10^{18}$ cm$^{-3}$, and $N_a = 10^{19}$ cm$^{-3}$ of P3HT to observe the variation in PCE, FF, $V_{OC}$, and $J_{SC}$ for the best absorber thickness. Their observed trends have been exhibited in figure 6.

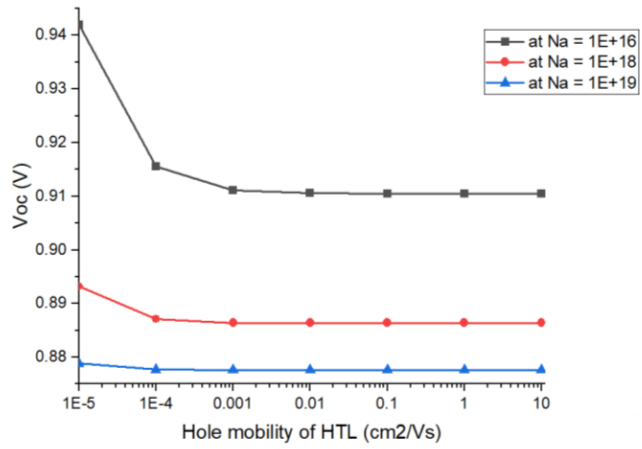

**(a)**

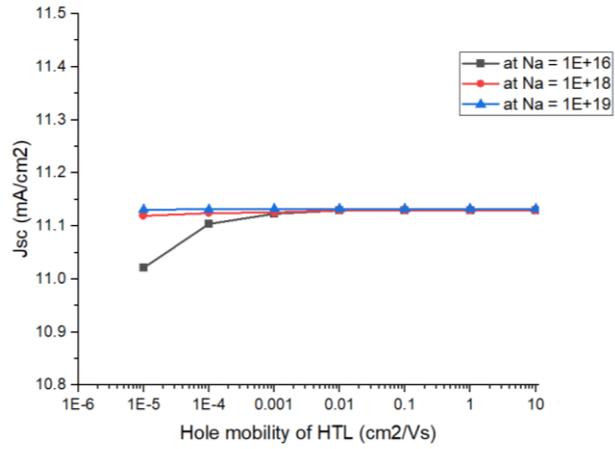

**(b)**

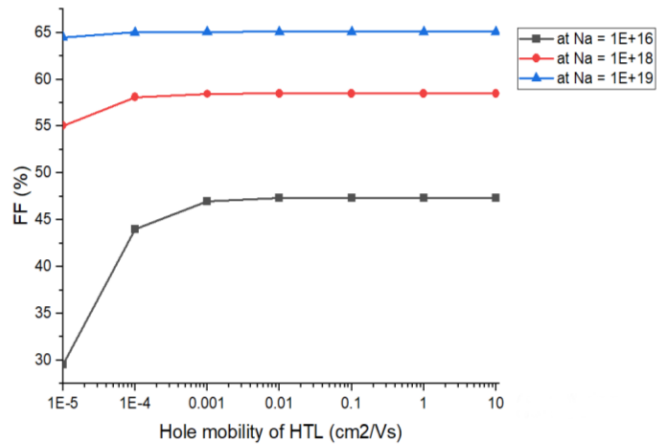

(c)

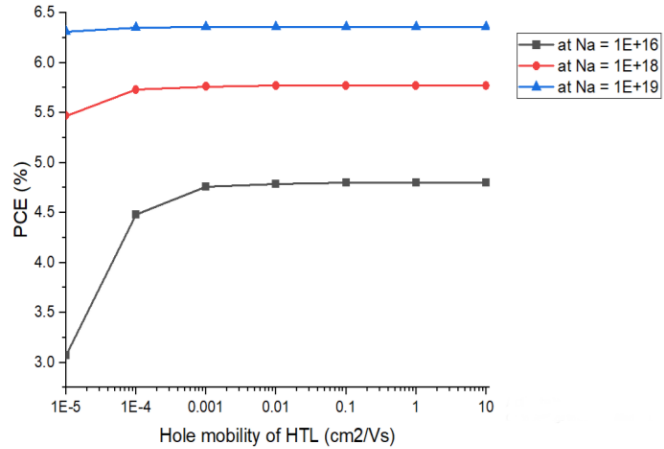

(d)

**Figure 6**: **(a)** $V_{OC}$, **(b)** $J_{SC}$, **(c)** FF, and **(d)** PCE diagram with varying $\mu_p$ of P3HT for $N_a = 10^{16}$ cm$^{-3}$, $N_a = 10^{18}$ cm$^{-3}$, and $N_a = 10^{19}$ cm$^{-3}$ of P3HT for the best absorber thickness.

The $\mu_p$ of HTL indicates the amount of hole transportation under the electric field's action and is affected by the $N_a$ of HTL. Lattice scattering restricts the $\mu_p$ at the lower $N_a$, but ionized impurity scattering limits the $\mu_p$ at the higher $N_a$.[79] With the increasing $\mu_p$ of P3HT HTL, the PCE, FF, and $J_{SC}$ of the PSC followed an increasing trend, whereas the $V_{OC}$ followed a downward trend. The trends were similar to the previously published literature[82] and were applicable for each $N_a$ of P3HT. With the increasing $\mu_p$ of P3HT, the PCE, $V_{OC}$, FF, and $J_{SC}$ became saturated after reaching a certain $\mu_p$ for each $N_a$ of P3HT, which was taken as their corresponding optimum $\mu_p$. At the base $N_a = 10^{16}$ cm$^{-3}$ of P3HT, the PCE reached to a maximum value of 4.8% ($J_{SC}$ = 11.13 mA/cm$^2$, FF = 47.35%, and $V_{OC}$ = 0.9105 V), when the $\mu_p$ was $10^{-1}$ cm$^2$V$^{-1}$s$^{-1}$, and so, it was taken as the optimum $\mu_p$. Similarly, for $N_a = 10^{18}$ cm$^{-3}$, the optimum $\mu_p$ was $10^{-2}$ cm$^2$V$^{-1}$s$^{-1}$ (PCE = 5.77%), and for $N_a = 10^{19}$ cm$^{-3}$, the optimum $\mu_p$ was $10^{-3}$ cm$^2$V$^{-1}$s$^{-1}$ (PCE = 6.36%). The improvement of $J_{SC}$ and PCE with the increasing $\mu_p$ of P3HT showcases the better charge transport and charge extraction at the P3HT/ perovskite interface at the higher $\mu_p$.[82] Besides, it was found that at the higher $N_a$ of P3HT, the $\mu_p$ modulation of P3HT had a marginal impact on performance improvement. When the $\mu_p$ of P3HT was augmented from $10^{-5}$ cm$^2$V$^{-1}$s$^{-1}$ to 10 cm$^2$V$^{-1}$s$^{-1}$, there was a 1.73% rise in PCE (in magnitude) for the base $N_a = 10^{16}$ cm$^{-3}$ of P3HT. However, for $N_a = 10^{18}$ cm$^{-3}$ and $N_a = 10^{19}$ cm$^{-3}$, the amount of improvement was reduced to 0.3% and 0.05% in magnitude, respectively. Furthermore, at the higher $N_a$, all the performance parameters became saturated more rapidly with the increasing $\mu_p$, and that's why the optimum $\mu_p$ became lower. As previously stated, the device performance can be drastically improved by doping as it enhances the $\mu_p$ and charge density, resulting in increased conductivity.[86,87] However, deep Coulomb traps can also be generated at the higher $N_a$ with extra charges that can decline the $\mu_p$. Therefore, the $\mu_p$ and $N_a$ of HTL should be weighed during the choosing of appropriate doping additives and doping levels.[82,88,89]

### 3.3 Effect of the Band-to-Band Radiative Recombination Rate

Band-to-band radiative recombination, Shockley-Read-Hall (SRH) recombination, and non-radiative Auger effect are the three generally considered recombination types for the PSC performance analysis. The total recombination rate for a solar cell can be expressed as a function of charge carrier density with the assistance of a cubic polynomial stated below:[90]

$$\frac{dn}{dt} = -k_3 n^3 - k_2 n^2 - k_1 n \ldots \ldots \ldots (5)$$

Where, *n* is the excess charge carrier density. The cubic, quadratic, and linear terms represent Auger, band-to-band radiative, and SRH recombination, respectively. The radiative recombination is contingent on the instantaneous recombination of electrons and holes generated from the electron-hole pairs that are produced during the absorption of light in the PSC. This phenomenon makes this recombination the most intrinsic form of recombination.[68] On the contrary, the SRH recombination is contingent on the defect states and their densities, while the Auger process becomes significant only at the high injection levels.[16,68,78] For the PSC, the band-to-band radiative recombination rate was augmented from the initial $2.3 \times 10^{-9}$ cm$^3$sec$^{-1}$ to $2.3 \times 10^{-8}$ cm$^3$sec$^{-1}$ as illustrated in figure 7 to observe the $V_{OC}$, $J_{SC}$, FF, and PCE trends of the PSC, and for each rate, the absorber thickness was varied from 200 nm to 900 nm. Also, the change in total recombination profile with different radiative recombination rates is shown in figure 8.

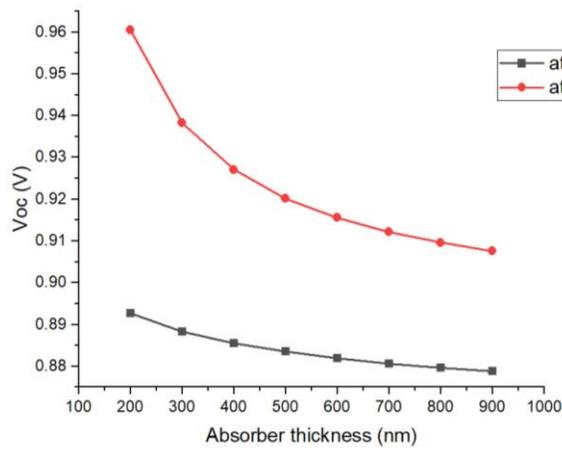

(a)

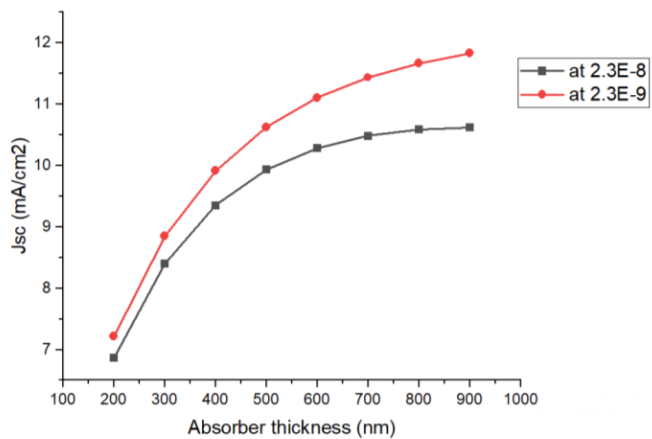

(b)

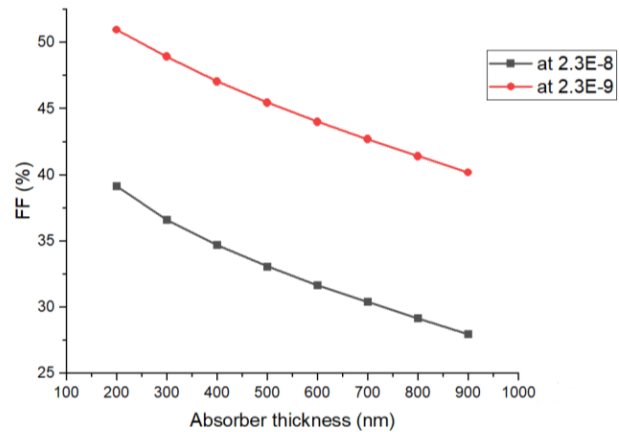

(c)

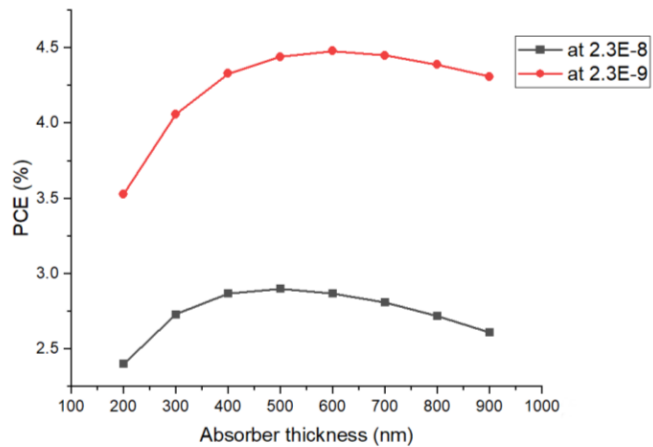

(d)

**Figure 7:** Solar performance outputs for the different band-to-band radiative recombination rates as a function of the absorber thickness. **a.** $V_{OC}$, **b.** $J_{SC}$, **c.** FF, and **d.** PCE

The $V_{OC}$, $J_{SC}$, FF, and PCE decreased at the higher radiative recombination rate according to figure 7, which was as expected. There is a substantial similarity between the trends of the parameters in figure 2 and figure 7 because of the radiative recombination's dependency on the perovskite layer's acceptor density ($N_a$) and donor density ($N_d$). This ascertains that radiative recombination is an inherent material property.[78] The $J_{SC}$ reduced at the higher radiative recombination rate indicating that the generation and transportation of electron-hole pair of the PSC were affected by it. Furthermore, the increased radiative recombination rate had minimal effect on $V_{OC}$ as there was approximately 0.04 V declination only for each absorber thickness level. However, the radiative recombination exerted a more significant effect on FF. The FF decreased significantly (a magnitude of 7.5% approximately at each thickness level) at the higher radiative recombination rate that can be explained by the less effective transportation of electron-hole pair through the PSC with more losses.[16,68,91] Finally, the PCE is a function of $V_{OC}$, $J_{SC}$, and FF. A radiative recombination rate of $2.3×10^{-9}$ $cm^3 sec^{-1}$ showed a maximum PCE of 4.48% at 600 nm, and a recombination rate of $2.3×10^{-8}$ $cm^3 sec^{-1}$ revealed a maximum PCE of 2.9% at 500 nm. This highlighted a significant PCE reduction of 1.58% in magnitude. This analysis suggests that the PCE of the PSC can be altered experimentally by changing the composition of $Cs_2AgBiBr_6$ through doping, thus influencing the radiative recombination.[78]

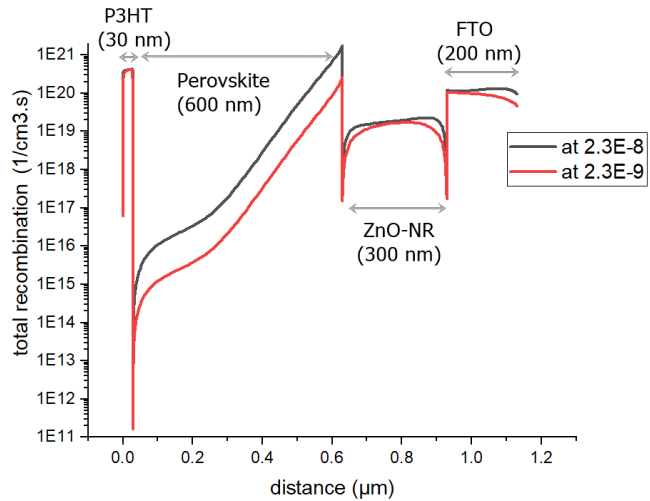

**Figure 8**: Variation in total recombination profile of the PSC with different radiative recombination rates for 600 nm absorber thickness.

## 3.4 Effect of the Electron Affinity (EA) of the HTL and ETL

The electron affinity (EA) of $Cs_2AgBiBr_6$ needs to be higher than the EA of HTL to ensure proper extraction of holes at the $Cs_2AgBiBr_6$/ HTL interface. Similarly, for the suitable extraction of electrons at the ETL/ $Cs_2AgBiBr_6$ interface, the EA of ETL needs to be higher than the EA of $Cs_2AgBiBr_6$. Figure 9 represents the energy level diagram for the PSC. To understand the impact of the EA of P3HT HTL on the PSC performance, the EA of P3HT was varied from 2.8 eV to 3.6 eV for the best absorber thickness (while keeping the EA of ZnO-NR fixed), and the observed trends in PCE, FF, $V_{OC}$, and $J_{SC}$ have been shown in figure 10. For the same reason, the EA of ZnO-NR was later modulated from 3.6 eV to 5.1 eV for the best absorber thickness (while keeping the EA of P3HT unchanged), and the variations in PCE, FF, $V_{OC}$, and $J_{SC}$ have been captured in figure 11. In both cases, the EA of $Cs_2AgBiBr_6$ and bandgap of each layer were kept constant. The $V_{OC}$ and $J_{SC}$ of the PSC were affected with varying EA due to the energy level mismatches at the perovskite/ HTL and ETL/ perovskite interfaces.

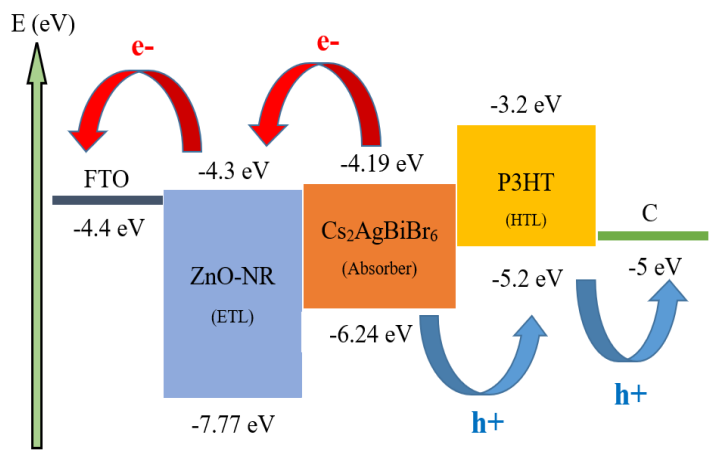

**Figure 9:** Energy level diagram for the PSC.

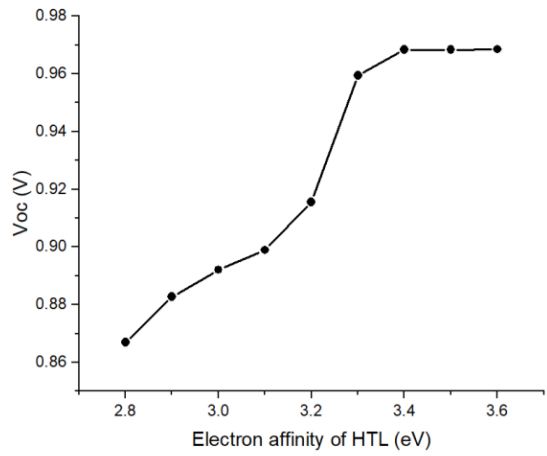

**(a)**

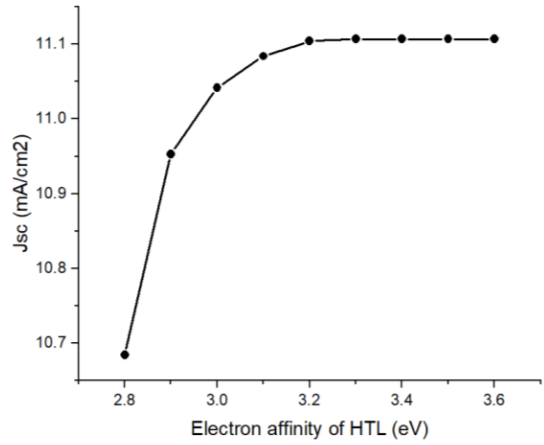

**(b)**

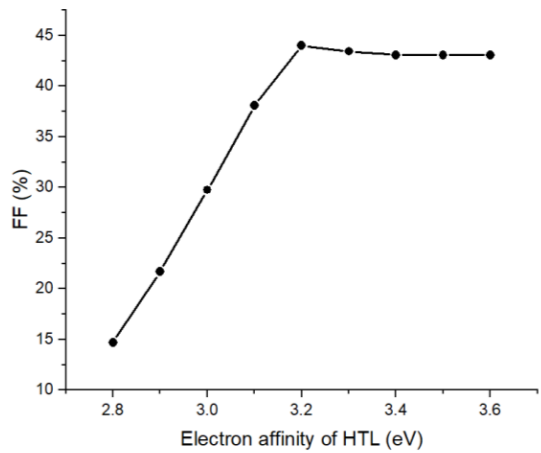

**(c)**

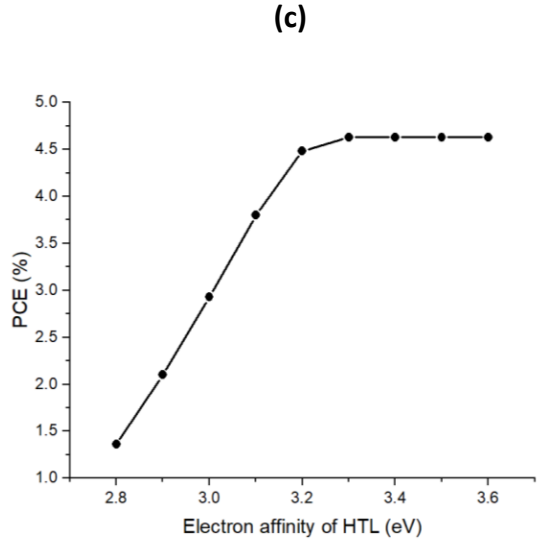

**(d)**

**Figure 10**: **(a)** $V_{OC}$, **(b)** $J_{SC}$, **(c)** FF, and **(d)** PCE diagram with varying EA of P3HT HTL for the best absorber thickness.

When the EA of HTL increases, the valence band offset (VBO) also increases as VBO = $(E_V)_{absorber}$ - $(E_V)_{HTL}$, where $E_V$ is the valence band. Initially, the VBO was -1.04 eV. As the EA of P3HT HTL was increased from 2.8 eV to 3.6 eV according to figure 10, the VBO also increased from -1.44 eV to -0.64 eV. The PCE, $J_{SC}$, and $V_{OC}$ followed similar trends to the published literature.[68] On the contrary, the trend in FF was almost similar to the literature.[78] When the EA of P3HT was enhanced from 2.8 eV to 3.4 eV, the $J_{SC}$ increased by 0.4222 mA/cm$^2$. After that, the $J_{SC}$ remained unchanged till 3.6 eV. Moreover, the $V_{OC}$ increased steadily when the EA of P3HT was increased from 2.8 eV to 3.6 eV with a rise of 0.1016 V. The activation energy ($E_a$) for carrier recombination at the perovskite/ HTL interface is shown by $E_{g, perovskite}$ - |VBO|. So, when the VBO decreased with the reducing EA of P3HT, the $E_a$ also got reduced. The interface recombination enhances with the decreasing $E_a$ that can cause a reduction in $V_{OC}$.[80,92] However, the change in FF was more significant than the change in $J_{SC}$ and $V_{OC}$. The FF increased from 14.71% to 44.02% in the range between 2.8 eV and 3.2 eV. But after 3.2 eV EA of P3HT, the FF started to decline. Finally, reaching a value of 43.07% at 3.5 eV EA of P3HT, the FF remained stable. As the PCE is a function of FF, $V_{OC}$, and $J_{SC}$, there was a rise of 3.27% in PCE (in magnitude) when the EA of P3HT was modulated from 2.8 eV to 3.3 eV. However, the PCE remained invariable later on. Therefore, at the 3.3 eV optimum EA of P3HT, the maximum PCE attained was 4.63%.

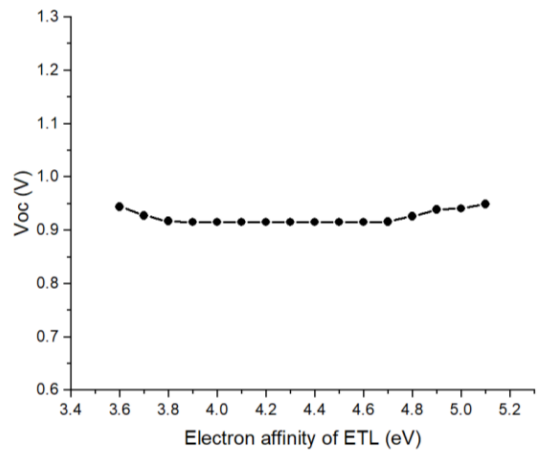

(a)

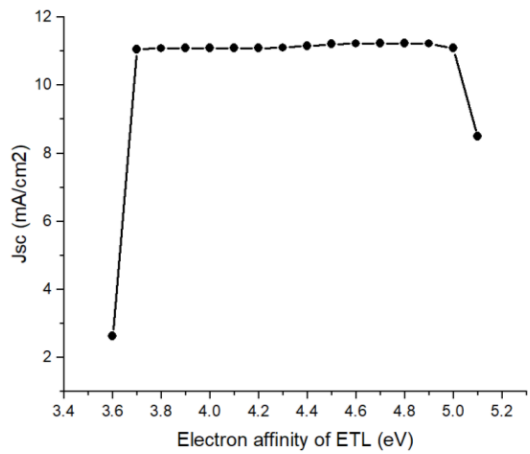

(b)

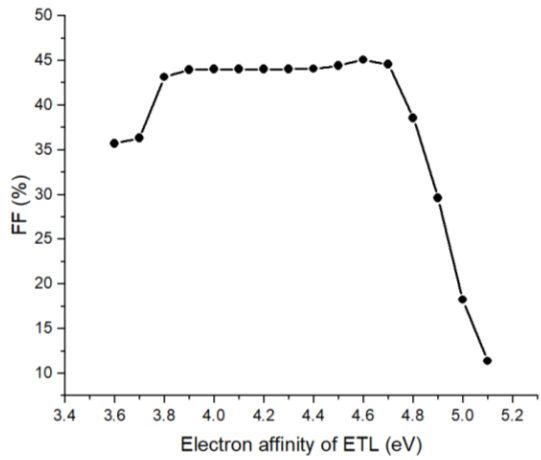

(c)

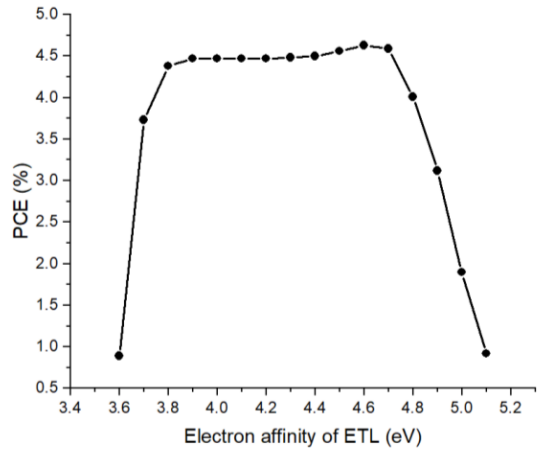

(d)

**Figure 11**: **(a)** $V_{OC}$, **(b)** $J_{SC}$, **(c)** FF, and **(d)** PCE diagram with varying EA of ZnO-NR ETL for the best absorber thickness.

Contrarily, with the increasing EA of ETL, the conduction band offset (CBO) decreases as CBO = $(E_C)_{ETL}$ - $(E_C)_{absorber}$. Here, $E_C$ is the conduction band. Therefore, primarily, the CBO was -0.11 eV, and when the EA of ZnO-NR ETL was increased from 3.6 eV to 5.1 eV according to figure 11, the CBO got reduced from 0.59 eV to -0.91 eV. The trends in PCE, FF, $J_{SC}$, and $V_{OC}$ were almost similar to the published literature.[93] The CBO at the absorber/ ETL interface regulates the transmission of carriers through the contact. In addition, the band adaptation in that interface changes from a 'spike' form to a 'cliff' form with the increasing EA of ETL.[92] The $V_{OC}$ remained almost invariable from 3.9 eV to 4.6 eV EA of ZnO-NR (or from 0.29 eV to -0.41 eV CBO). But the $V_{OC}$ got slightly improved outside that range, which could be due to the creation of a barrier from the ETL to the perovskite layer.[65,93] The PCE, FF, and $J_{SC}$ increased to a maximum value with the increasing EA of ZnO-NR (or the decreasing CBO). After reaching the maximum value, they started to decline. When the EA of ZnO-NR was modulated from 3.7 eV to 5 eV, the fluctuation in $J_{SC}$ was less pronounced. However, outside of that range, there was a significant reduction in $J_{SC}$. Similarly, the reduction in PCE and FF was impactful outside the range of 3.8 eV to 4.7 eV EA of ZnO-NR; otherwise, the change was very marginal within the range. The maximum PCE of 4.63% along with the maximum FF of 45.08% (with $V_{OC}$ = 0.9153 V and $J_{SC}$ = 11.2259 mA/cm$^2$) was obtained at 4.6 eV EA of ZnO-NR. When the CBO is too high or too low, the probability of recombination in the PSC increases. This causes a drastic drop in PCE, FF, and $J_{SC}$. Here, the reduction in the photogenerated carrier density can be influenced by the SRH, radiative, Auger, and grain boundary or surface recombination.[93]

**3.5 Effect of the Back Contact Work Function**

In this simulation study, we initially took C as the back contact with a work function of 5 eV. Meanwhile, Ag, Cr, Cu, Fe, Au, and Ni have been commonly employed as the back contact of PSC. Their corresponding work functions are 4.26 eV, 4.5 eV, 4.65 eV, 4.74 eV, 5.1 eV, and 5.15 eV, respectively.[8,71,94] The back contact work function of the PSC was modulated from 4.2 eV to 5.4 eV for the best absorber thickness to investigate its influence on the PSC performance. The observed trends in PCE, FF, $V_{OC}$, and $J_{SC}$ have been illustrated in figure 12.

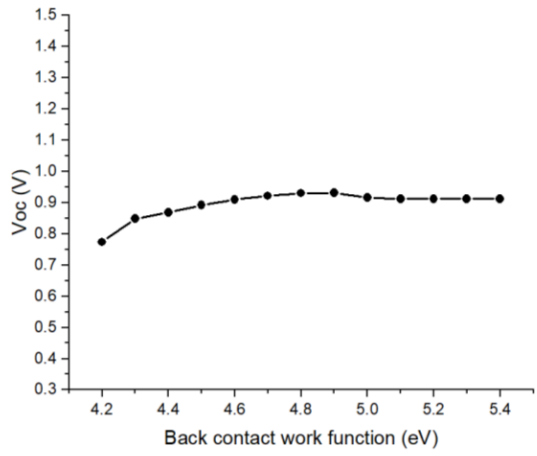

**(a)**

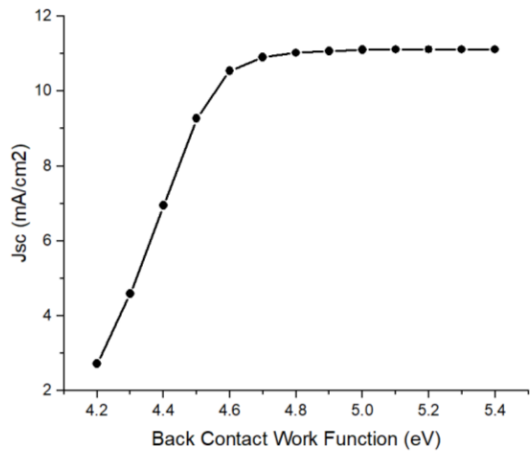

**(b)**

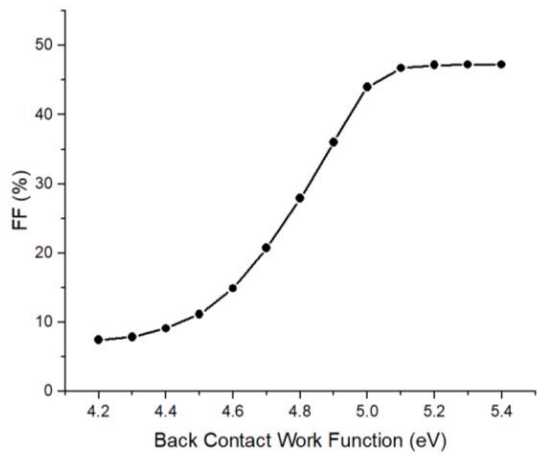

**(c)**

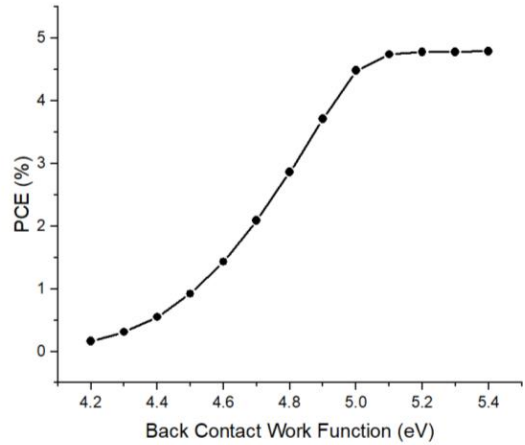

**(d)**

**Figure 12**: **(a)** $V_{OC}$, **(b)** $J_{SC}$, **(c)** FF, and **(d)** PCE diagram with the varying work function of the back contact for the best absorber thickness.

The tendency for electrons to tunnel through the HTL and enter the back contact declines with the increasing back contact work function, whereas the holes have a higher driving force to enter the back contact from the HTL.[78] With the increasing back contact work function from 4.2 eV to 4.9 eV, the $V_{OC}$ increased steadily. However, after the 4.9 eV work function, there was a slight drop in $V_{OC}$. After that, the $V_{OC}$ remained almost invariable later on. Meanwhile, increasing the work function enhanced the $J_{SC}$, FF, and PCE steadily up to 5.2 eV. But after 5.2 eV, they became almost stable. So, the optimum back work function was 5.2 eV with a PCE of 4.78% (with $V_{OC}$ = 0.9117 V, $J_{SC}$ = 11.1134 mA/cm$^2$, and FF = 47.16%). Therefore, Ni can be a suitable replacement for C in the PSC to improve the device's performance. Because Ni has a work function of 5.15 eV, which is near the optimum value. The Fermi level energy reduces at the higher back contact work function due to the band bending at the metal-semiconductor interface, which assists the contact to be more ohmic.[8,95] The built-in voltage in a PSC diminishes when the back contact work function is low that causes a drop in $V_{OC}$. Simultaneously, an inefficient collection of photogenerated carriers occurs, dropping the $J_{SC}$.[96] When the back contact work function was below 5.1 eV, there was a drastic declination in the PSC performance. This could be due to the formation of a Schottky junction at the HTL/ back contact interface that can enhance the series resistance.[8,97,98] As the back contact work function reduces, the Schottky barrier enhances. This restricts the hole transportation, causing the FF and PCE to diminish.[96]

**3.6 Effect of the Defect Density ($N_t$) of the Absorber Layer**

As previously stated, all of the layers were considered defect-free this far to get the theoretical maximum of the device performance that can be approached from an experimental point of view. But in reality, there are various defects in each layer, which restricts the performance of the PSC depending on their severity. The properties of each layer's defects are shown in table 2 that were taken cautiously from the published literature.[65,68,73] Experimental value suggests that the $Cs_2AgBiBr_6$ perovskite has a defect density ($N_t$) of $9.1 \times 10^{16}$ cm$^{-3}$.[27] Using the advanced first-principle calculations, investigations have been carried out to probe the nature of defects in the $Cs_2AgBiBr_6$ structure.[99] These intrinsic point defects are vacancies, interstitials, and antisite defects, and investigations have been performed to correlate them to their experimental growth conditions and chemical potentials.[72] Moreover, due to the weak photoluminescence intensity measured for $Cs_2AgBiBr_6$ material, its dominant recombination mechanism is nonradiative.[100] The electrons and holes generated by the absorption of photons in the absorber are separated and transported to the ETL and HTL in the built-in electric field, respectively. So, the recombination through bulk defects in the absorber is a crucial mechanism influencing the performance of the PSC.[80] The absorber $N_t$ was varied between $10^{14}$ cm$^{-3}$ to $10^{20}$ cm$^{-3}$ in the simulation to find the variation in PCE, FF, $V_{OC}$, and $J_{SC}$ for the best absorber thickness. The simulated results have been plotted in figure 13. In addition, the corresponding QE vs. wavelength curve and the alteration in total recombination profile of the PSC with different absorber $N_t$ have been shown in figure 14.

**Table 2**: Properties of the defects in each layer used in the simulation.

| Layers | Defect type | Defect Density | Capture cross-section electrons (cm²) | Capture cross-section holes (cm²) | Energy level w.r.t. reference (eV) |
|---|---|---|---|---|---|
| P3HT (HTL) | Neutral | 1.0E+15 | 1.0E-15 | 1.0E-15 | 0.60 (above EV) |
| $Cs_2AgBiBr_6$ (Absorber) | Neutral | Variable | 1.0E-15 | 1.0E-14 | 0.05 (above EV) |
| ZnO-NR (ETL) | Neutral | 1.0E+15 | 2.0E-14 | 2.0E-14 | 0.60 (above EV) |
| FTO (TCO) | Neutral | 1.0E+15 | 1.0E-15 | 1.0E-15 | 0.60 (above EV) |

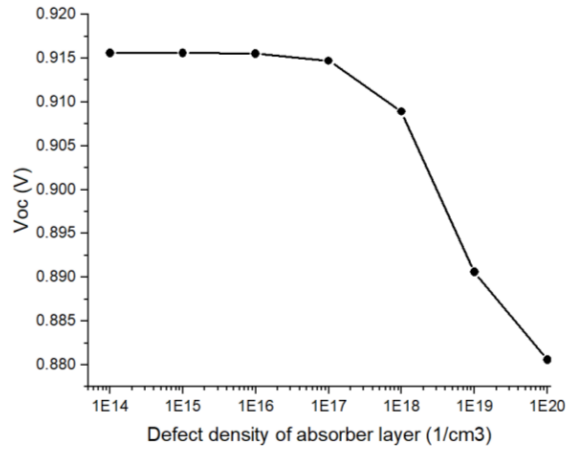

(a)

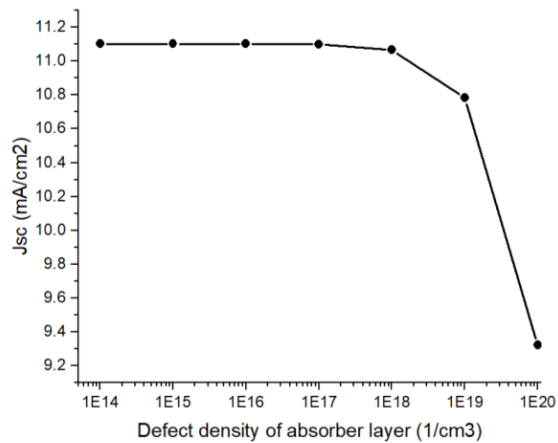

(b)

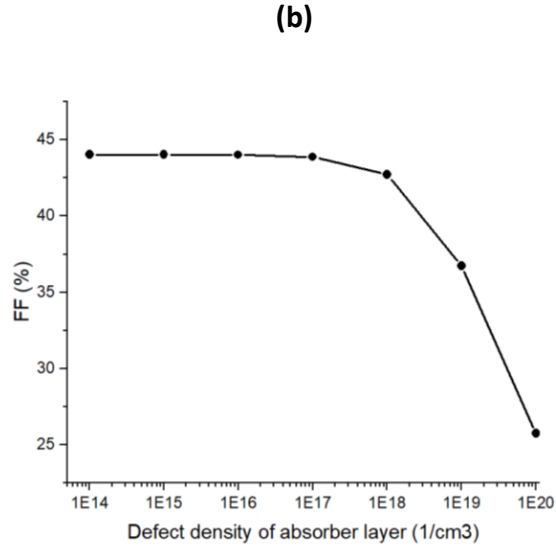

(c)

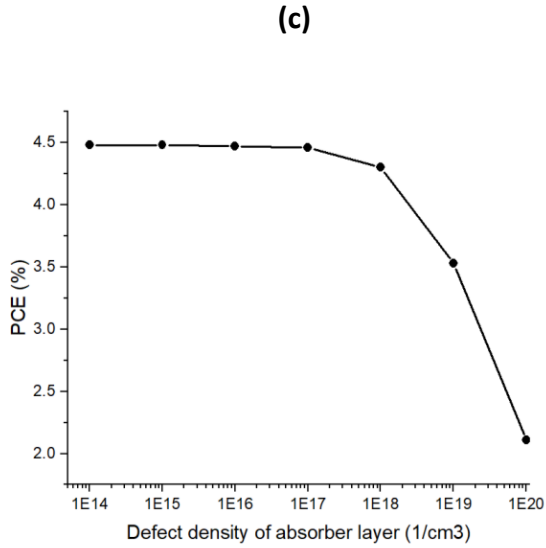

(d)

**Figure 13**: **(a)** $V_{OC}$, **(b)** $J_{SC}$, **(c)** FF, and **(d)** PCE diagram with varying absorber defect density ($N_t$) for the best absorber layer thickness.

As the absorber $N_t$ was reduced from $10^{20}$ cm$^{-3}$ to $10^{15}$ cm$^{-3}$, the PCE, FF, $V_{OC}$, and $J_{SC}$ of the PSC improved. During this modulation, the PCE faced a significant improvement of 2.37% in magnitude. However, all the performance parameters became saturated when the $N_t$ was lower than $10^{15}$ cm$^{-3}$. Therefore, $10^{15}$ cm$^{-3}$ defect density was determined to be the optimum absorber $N_t$. At this optimum value, the maximum PCE of 4.48% (with $J_{SC}$ = 11.1027 mA/cm², FF = 44.02%, and $V_{OC}$ = 0.9156 V) was observed, which was equal to the defect-free PCE. The defect states in the perovskite can introduce additional carrier recombination centers that facilitate the recombination process of the photo-excited carriers.[101] As a result, the reverse saturation current enhances, and parameters like the diffusion length, $V_{OC}$, $J_{SC}$, and PCE deteriorate.[102] In addition, the electrons in transition between the bands pass through a localized energy state created by a dopant or defect within the bandgap.[103] This localized trap is a potential site where the non-radiative recombination primarily takes place. This trap-supported recombination is called the SRH recombination as stated previously. It is a commanding cause of carrier recombination, lifetime reduction, and radical drop in the device performance.[103,104] Furthermore, to quantitatively analyze the influence of defect states on the PSC performance, the Gaussian distributions can be incorporated in the perovskite layer.[17,105] The corresponding equations of SRH recombination and Gaussian distributions are provided in the Supplementary File.

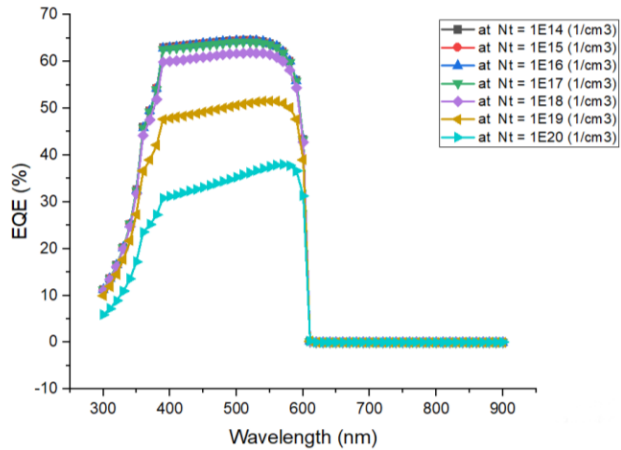

(a)

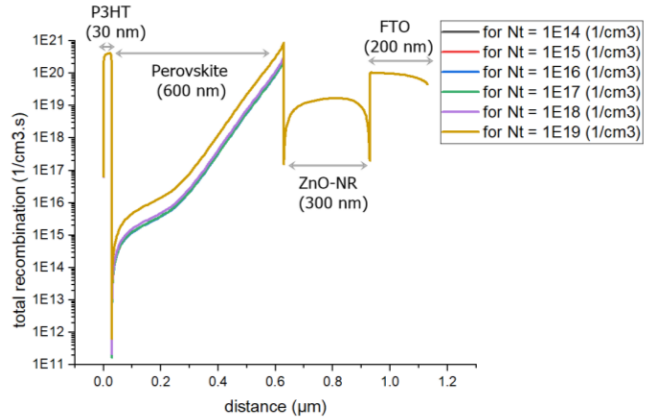

(b)

**Figure 14**: **(a)** QE vs. wavelength diagram and **(b)** total recombination profile diagram for different absorber defect densities.

### 3.7 Study of the Alternative Charge Transport Layers

We optimized the absorber defect density of SLG/ FTO/ ZnO-NR/ $Cs_2AgBiBr_6$/ P3HT/ C solar cell for the best absorber thickness previously. But several other materials have been employed as the charge transport layers in the PSCs. Therefore, we replaced the ZnO-NR ETL with inorganic $TiO_2$, $SnO_2$, CdS, and organic phenyl-C61-butyric acid methyl ester (PCBM) as well as P3HT HTL with inorganic $Cu_2O$, CuO, $MoO_3$, and organic PEDOT: PSS in the simulation to investigate the modulation in the device performance. The material parameters utilized for the alternative charge transport layers in the simulation have been provided in table 3.[68,73,85,106-110] Furthermore, we used the optimum absorber thickness and absorber $N_t$ in the simulation. For proper comparison, we also kept the defect properties and doping concentration of alternative ETLs and HTLs similar to the base ZnO-NR ETL and P3HT HTL, respectively.

**Table 3**: Input material parameters used for the alternative ETLs and HTLs in the PSC simulation.

|  | Alternative ETLs | Alternative HTLs |
|---|---|---|

| Material | TiO$_2$ | SnO$_2$ | CdS | PCBM | Cu$_2$O | PEDOT: PSS | MoO$_3$ | CuO |
|---|---|---|---|---|---|---|---|---|
| E$_g$ (eV) | 3.20 | 3.50 | 2.40 | 2.00 | 2.17 | 1.60 | 3.00 | 2.10 |
| χ (eV) | 4.10 | 4.00 | 4.18 | 3.90 | 3.20 | 3.40 | 2.50 | 3.20 |
| ε$_r$ | 9.00 | 9.00 | 10.00 | 3.90 | 7.10 | 3.00 | 12.50 | 7.11 |
| N$_c$ (cm$^{-3}$) | 2.2E+18 | 2.2E+17 | 2.2E+18 | 2.5E+21 | 2.5E+18 | 1.0E+22 | 2.2E+18 | 2.2E+20 |
| N$_v$ (cm$^{-3}$) | 1.0E+19 | 2.2E+16 | 1.9E+19 | 2.5E+21 | 1.8E+19 | 1.0E+22 | 1.8E+19 | 5.5E+19 |
| μ$_n$ (cm$^2$V$^{-1}$s$^{-1}$) | 20 | 20 | 100 | 0.02 | 200 | 4.5E-4 | 100 | 3.4 |
| μ$_p$ (cm$^2$V$^{-1}$s$^{-1}$) | 10 | 10 | 25 | 0.02 | 80 | 9.9E-5 | 25 | 3.4 |

Initially, we kept the P3HT HTL fixed and ran the simulation for different ETLs, and the device performance achieved for different configurations has been shown in table 4. Also, the QE vs. wavelength curves for different ETLs have been shown in figure 15 (a). It was found that the base ZnO-NR ETL showed the best PCE that was followed by SnO$_2$, TiO$_2$, CdS, and PCBM. However, there were only a 0.01% difference in PCE between ZnO-NR and SnO$_2$ and 0.07% between ZnO-NR and TiO$_2$ as the ETL (in magnitude). This suggests that SnO$_2$ and TiO$_2$ can be utilized as the replacement of ZnO-NR ETL. Moreover, it can be deduced from Table 4 and figure 15 (a) that SnO$_2$ and ZnO-NR as the ETL showed the best J$_{SC}$ because of the enhanced light absorption by the perovskite, which is also reflected on their overlapping QE curves.[95] They possessed wider bandgaps (3.5 eV for SnO$_2$ and 3.47 eV for ZnO-NR) and had consistently good light transmittance over the whole simulated wavelength range compared to the rest of the ETLs.

**Table 4**: Device performance for different ETLs with P3HT HTL in the descending order of PCE.

| Device | $V_{OC}$ (V) | $J_{SC}$ (mA/cm$^2$) | FF (%) | PCE (%) |
|---|---|---|---|---|
| ZnO-NR/ Cs$_2$AgBiBr$_6$/ P3HT (Base) | 0.9156 | 11.1027 | 44.02 | 4.48 |
| SnO$_2$/ Cs$_2$AgBiBr$_6$/ P3HT | 0.9156 | 11.0887 | 44.01 | 4.47 |
| TiO$_2$/ Cs2AgBiBr6/ P3HT | 0.9155 | 10.9002 | 44.09 | 4.40 |
| CdS/ Cs2AgBiBr6/ P3HT | 0.9140 | 7.4470 | 45.35 | 3.09 |
| PCBM/ Cs2AgBiBr6/ P3HT | 0.9111 | 2.7575 | 46.92 | 1.18 |

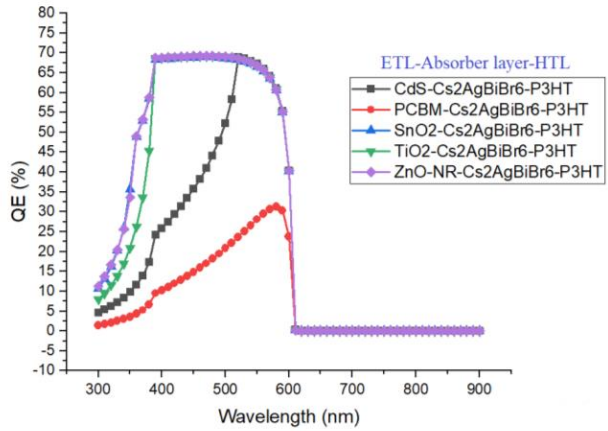

(a)

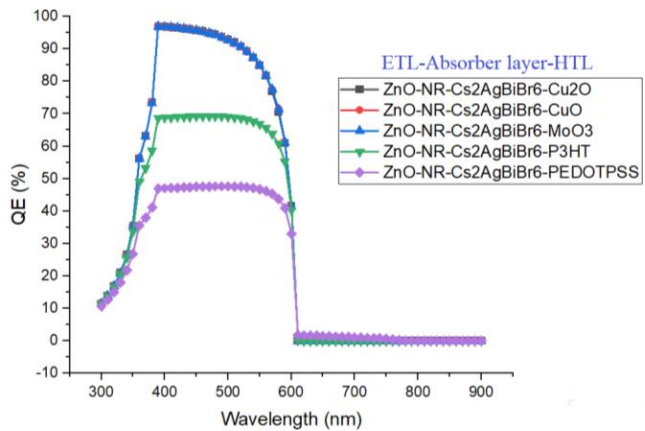

(b)

**Figure 15.** QE vs. wavelength diagram for (a) different ETLs and (b) different HTLs in the PSC.

Next, we kept the ZnO-NR ETL fixed in the PSC as it showed the best performance and ran the simulation for different HTLs. The device performance obtained for different configurations has been illustrated in table 5. Besides, the QE vs. wavelength curves for different HTLs have been illustrated in figure 15 (b). We could observe that the inorganic $Cu_2O$ HTL showed the best PCE, followed by CuO, $MoO_3$, P3HT, and PEDOT: PSS. Also, the PCE of the device with CuO and $MoO_3$ as the HTL had only 0.01% and 0.1% less PCE (in magnitude), respectively, than the device with $Cu_2O$. Therefore, CuO and $MoO_3$ can also be used as the alternative HTL in the PSC. As the HTL thickness (30 nm) was much smaller than the absorber thickness (600 nm) and it was positioned under the absorber layer, the HTL absorbed a very small fraction of the incident solar radiation. Therefore, the optical absorption coefficient of the HTL will be much smaller compared to the absorber layer. That's why it was observed from table 5 that the $J_{SC}$ for different HTL configurations was marginally impacted.[111-114] For the same reason, the QE curves for inorganic $Cu_2O$, CuO, and $MoO_3$ as the HTL overlapped as expected according to figure 15 (b), except for organic P3HT and PEDOT: PSS. This type of exception was previously reported in the published literature.[115] This was because, although the HTL will not significantly affect the electron-hole photogeneration, it can affect the photogenerated charge collection efficiency due to the energy level mismatches at the perovskite/ HTL interfaces.[111-114] Finally, we could conclude that a fully inorganic PSC with a structure of FTO/ ZnO-NR/ $Cs_2AgBiBr_6$/ $Cu_2O$/ C had the best PCE of 5.16%, which was 0.68% higher than the base structure in magnitude.

**Table 5**: Device performance for different HTLs with ZnO-NR ETL in the descending order of PCE.

| Device | $V_{OC}$ (V) | $J_{SC}$ (mA/cm$^2$) | FF (%) | PCE (%) |
|---|---|---|---|---|
| ZnO-NR/ $Cs_2AgBiBr_6$/ $Cu_2O$ | 1.0517 | 11.1681 | 43.97 | 5.16 |
| ZnO-NR/ $Cs_2AgBiBr_6$/ CuO | 0.9724 | 11.1663 | 47.43 | 5.15 |
| ZnO-NR/ $Cs_2AgBiBr_6$/ $MoO_3$ | 1.0045 | 11.1093 | 45.31 | 5.06 |
| ZnO-NR/ $Cs_2AgBiBr_6$/ P3HT (Base) | 0.9156 | 11.1027 | 44.02 | 4.48 |
| ZnO-NR/ $Cs_2AgBiBr_6$/ PEDOT: PSS | 0.8656 | 11.0998 | 20.54 | 1.97 |

## 4. Conclusion

In this work, SCAPS 1-D was utilized to carry out the simulation studies into a non-toxic, organic-inorganic double halide PSC using the $Cs_2AgBiBr_6$ as the absorber layer, P3HT as the HTL, ZnO-NR as the ETL, and C as the back contact. Upon investigation, the maximum PCE of 4.48% was observed at 600 nm optimum absorber thickness. Again, as the $N_a$ of P3HT was augmented from $10^{15}$ cm$^{-3}$ to $10^{19}$ cm$^{-3}$, the PCE improved steadily from 4.37% to 6.35%. Moreover, the $\mu_p$ of P3HT was enhanced from $10^{-5}$ cm$^2$V$^{-1}$s$^{-1}$ to $10$ cm$^2$V$^{-1}$s$^{-1}$ for $10^{16}$ cm$^{-3}$, $10^{18}$ cm$^{-3}$, and $10^{19}$ cm$^{-3}$ $N_a$ of P3HT. The optimum hole mobilities for each $N_a$ of P3HT were $10^{-1}$ cm$^2$V$^{-1}$s$^{-1}$ (PCE = 4.8%), $10^{-2}$ cm$^2$V$^{-1}$s$^{-1}$ (PCE = 5.77%), and $10^{-3}$ cm$^2$V$^{-1}$s$^{-1}$ (PCE = 6.36%), respectively. Also, it was apparent that the $\mu_p$ of P3HT exerted a greater influence on the PSC's performance at lower $N_a$ of P3HT. In addition, the EA of P3HT was modulated from 2.8 eV to 3.6 eV, and the optimum EA was 3.3 eV, achieving a PCE of 4.63%. Similarly, as the EA of ZnO-NR was changed from 3.6 eV to 5.1 eV, the maximum PCE of 4.63% was found at 4.6 eV EA. Furthermore, there was a significant reduction in maximum PCE by 1.58% in magnitude when the radiative recombination rate was increased by 10 times. Besides, the back contact work function modulation from 4.2 eV to 5.4 eV suggested that the optimum work function was 5.2 eV with a PCE of 4.78%. Therefore, replacing C with Ni as the back contact will improve the device's performance. When the defects were introduced in each layer, the absorber's $N_t$ was varied from $10^{14}$ cm$^{-3}$ to $10^{20}$ cm$^{-3}$. The most favorable PCE value of 4.48% at $10^{15}$ cm$^{-3}$ optimum $N_t$ was observed. Meanwhile, the performance analysis using different charge transport layers suggested that ZnO-NR ETL showed the best performance and replacing P3HT with $Cu_2O$ as the HTL increased the PCE by 0.68% in magnitude. Finally, wxAMPS was employed for the absorber thickness optimization study for validation purposes. Overall, the insightful simulated results presented in this work will go a long way in providing recommendations towards experimentally realizing a viable, non-toxic PSC.

## 5. Acknowledgement

The authors acknowledge Dr. Marc Burgelman at the University of Gent, Belgium, for providing the simulation software SCAPS 1-D. They also acknowledge Prof. A. Rockett and Dr. Yiming Liu from UIUC and Prof. Fonash of PSU for providing the wxAMPS program.

# Supplementary File

**Computational equations utilized in wxAMPS:**

In 1-D space, Poisson's equation is given by:

$$\frac{d}{dx}\left(-\varepsilon(x)\frac{d\psi'}{dx}\right) = q.[p(x) - n(x) + N_D^+(x) - N_A^-(x) + pt(x) - nt(x)] \ldots \ldots \ldots (1)$$

Where, the electrostatic potential $\psi'$ and the free electron n, free hole p, trapped electron nt, and trapped hole pt as well as the ionized donor-like doping $N_D^+$ and ionized acceptor-like doping $N_A^-$ concentrations are all functions of the position coordinate x. The continuity equation for the free electrons in the delocalized states of the conduction band has the form:

$$\frac{1}{q}\left(\frac{dJn}{dx}\right) = -G_{op}(x) + Rx \ldots \ldots \ldots (2)$$

Again, the continuity equation for the free holes in the delocalized states of the valence band has the form:

$$\frac{1}{q}\left(\frac{dJp}{dx}\right) = G_{op}(x) - Rx \ldots \ldots \ldots (3)$$

Where, Jn and Jp are, respectively, the electron and hole current densities. The term R(x) is the net recombination rate resulting from band-to-band (direct) recombination and SRH (indirect) recombination traffic through gap states. The net direct recombination rate is:

$$R_D(x) = \beta(np - n_i^2) \ldots \ldots \ldots (4)$$

Where, $\beta$ is a proportionality constant, which depends on the material's energy band structure under analysis, and n and p are the band carrier concentrations present when devices are subjected to a voltage bias, light bias, or both. The continuity equations include the term $G_{op}(x)$, which is the optical generation rate as a function of x due to externally imposed illumination.

## Absorption data for the perovskite solar cell:

Absorption data for each layer was achieved from the new Eg-sqrt model (SCAPS version 3.3.07), which is the updated model of the traditional SCAPS model (traditional sqrt ($h\upsilon$-$E_g$) law model) and can be found from the "Tauc laws". The updated Eg-sqrt model follows equation 5.

$$\alpha(h\upsilon) = (\alpha_0 + \beta_0 \frac{E_g}{h\upsilon}) \sqrt{\frac{h\upsilon}{E_g} - 1} \ldots \ldots \ldots (5)$$

Where, $\alpha$ is the optical absorption constant, $h\upsilon$ is the photon energy, and $E_g$ is the bandgap. The model constants $\alpha_0$ and $\beta_0$ have the dimension of absorption constant (e.g., 1/cm) and are related to the traditional model constants A and B by the relations:

$$\alpha_0 = A\sqrt{E_g} \text{ and } \beta_0 = \frac{B}{\sqrt{E_g}}$$

## Gaussian Distribution and SRH Recombination:

The corresponding equations of the Gaussian distribution model that incorporates defect density are:

$$g_D(E) = G_{Md} \exp[\frac{-(E-E_{pkd})^2}{2\sigma_d^2}] \ldots \ldots \ldots (6)$$

$$g_A(E) = G_{Ma} \exp[\frac{-(E-E_{pka})^2}{2\sigma_a^2}] \ldots \ldots \ldots (7)$$

Where, $G_{Md}$ and $G_{Ma}$ are the effective defect densities, $\sigma_d$ and $\sigma_a$ are the standard energy deviations of the Gaussian donor and acceptor levels, $E_{pkd}$ and $E_{pka}$ are the donor peak energy position measured positive from $E_C$, and the acceptor peak energy position measured positive from $E_V$. The charge carriers in perovskite solar cell are recombined by Shockley-Read-Hall (SRH) recombination process and the net recombination rate ($R^{SRH}$) for SRH recombination is given by the following equation:

$$R^{SRH} = \frac{v\sigma_n\sigma_p N_T [np - n_i^2]}{\sigma_p[p + p_1] + \sigma_n[n + n_1]} \quad \ldots \ldots \ldots (8)$$

Where, $\sigma_n$ and $\sigma_p$ are the capture cross-sections for electrons and holes, $v$ is the electron thermal velocity, $N_T$ is the atomistic defect concentration, $n_i$ is the intrinsic carrier density, $n$ and $p$ are the concentrations of electron and hole at equilibrium, and $n_1$ and $p_1$ are the concentrations of electrons and holes in trap defect and valence band, respectively. According to equation 8, $R^{SRH}$ is directly proportional to the defect density in the perovskite absorber layer. Again, $R^{SRH}$ has an impact on the carrier diffusion length. The diffusion length increases with decreasing the perovskite absorber layer's defect density, which improves the solar cell performance. The relation between the diffusion length, carrier mobility, and lifetime at a temperature T is expressed in equation 9.

$$L_D = \sqrt{\frac{\mu_{(e,h)} R^{SRH} T}{q} \times \tau_{lifetime}} \quad \ldots \ldots \ldots (9)$$

Where, $L_D$, $\mu_{(e,h)}$, and $\tau_{lifetime}$ are the diffusion length, the electron and hole mobility, and the minority-carrier lifetime, respectively. Moreover, $\tau_{lifetime}$ depends upon the defect density and capture cross-section area for electrons and holes. The relation between $\tau_{lifetime}$ and bulk defect density is expressed in equation 10.

$$\tau_{lifetime} = \frac{1}{N_T \delta v_{th}} \ldots \ldots \ldots (10)$$

Here, $\delta$, $v_{th}$, and $N_T$ represent the capture cross-section area for electrons and holes, the thermal velocity of carriers, and defect concentration, respectively.

**Validation of the model using wxAMPS:**

The simulations were run using wxAMPS (version 2.0) with the aim to validate the results accumulated by the SCAPS software platform. The absorber layer thickness was varied between 150 nm to 900 nm in both software to find variation in $V_{OC}$, $J_{SC}$, FF, and PCE of the perovskite solar cell. All the simulations were performed at 300 K working temperature and AM1.5G solar spectrum. Table 1 shows the comparison between the two simulation tools in $V_{OC}$, $J_{SC}$, FF, and PCE at the optimum absorber thickness. Figure 1 shows the comparison between the SCAPS and wxAMPS results during the absorber layer thickness optimization study.

**Table 1.** Comparison between the SCAPS and wxAMPS results at the optimum absorber thickness.

| Serial | Software | Optimum Absorber Thickness (nm) | $V_{OC}$ V | $J_{SC}$ mA/cm2 | Fill Factor % | Efficiency % |
|---|---|---|---|---|---|---|
| 1 | SCAPS | 600 | 0.9156 | 11.104 | 44.02 | 4.48 |
| 2 | wxAMPS | 250 | 1.4008 | 11.2562 | 31.51 | 4.97 |

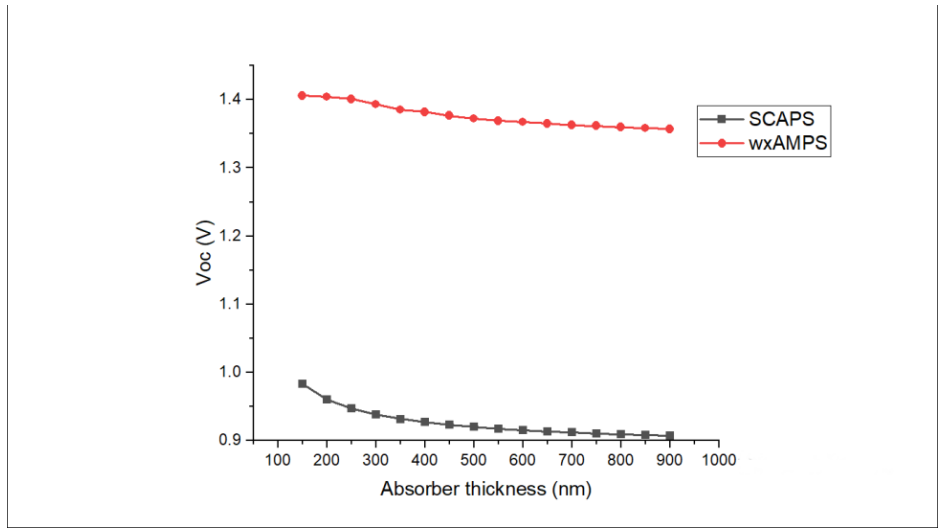

(a)

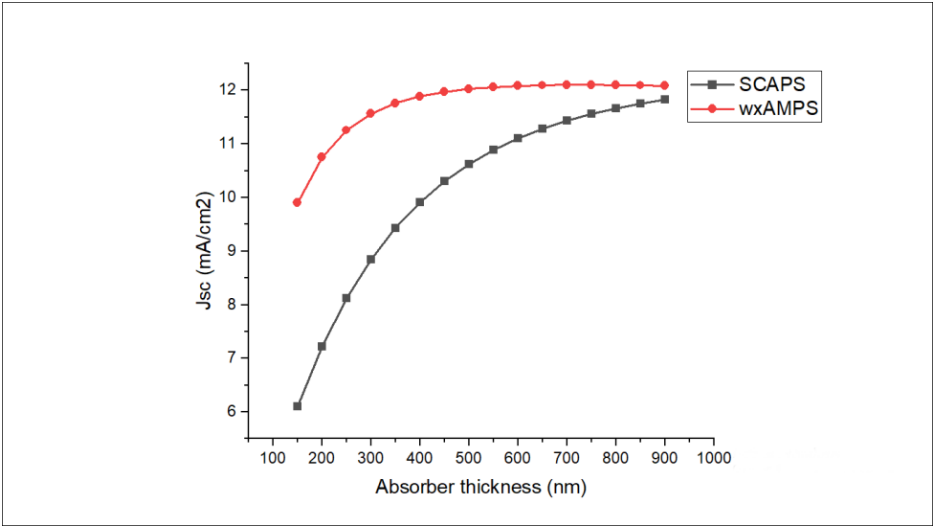

**(b)**

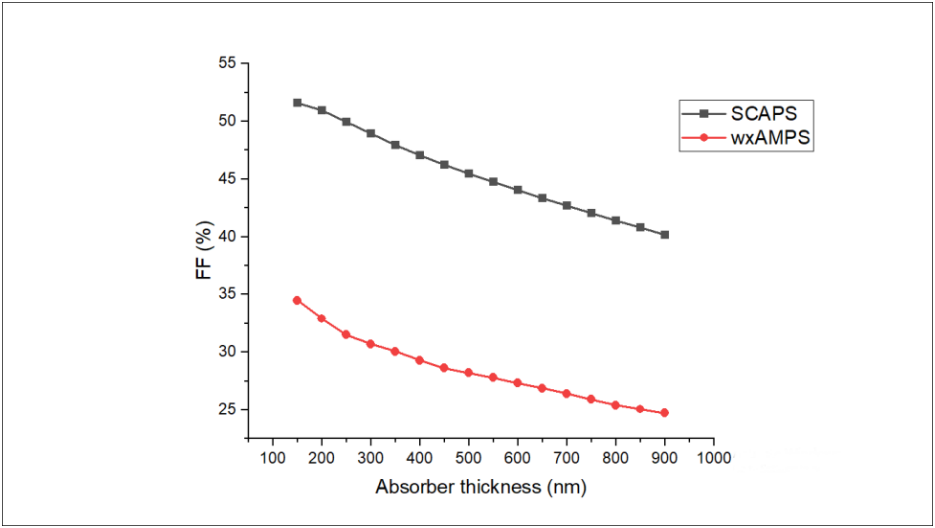

**(c)**

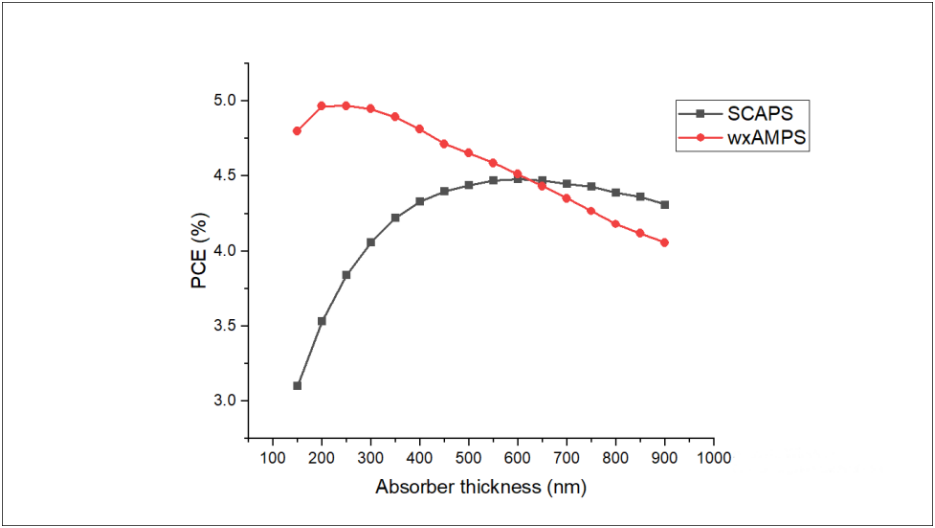

**(d)**

**Figure 1:** Comparison between the wxAMPS and SCAPS results in the absorber layer thickness optimization study.